\documentclass[aip,sd,amsmath,amssymb,preprint,graphicx]{revtex4-1}
\usepackage{graphicx}
\usepackage{dcolumn}
\usepackage{bm}

\draft 

\begin{document}


\title{Mass Density Fluctuations in Quantum and Classical descriptions of Liquid Water} 



\author{Mirza Galib}
\email{mirza.galib@pnnl.gov}
\affiliation
{Physical Sciences Division, Pacific Northwest National Laboratory, Richland, Washington 99354, United States}
\author{Timothy T. Duignan}
\author{Yannick Misteli}
 \altaffiliation[]{University of Zurich, Department of Chemistry, CH-8057,
 Switzerland}
\author{Marcel D. Baer}
\author{Gregory K. Schenter}
\author{J\"urg Hutter}
 \altaffiliation[]{University of Zurich, Department of Chemistry, CH-8057,
 Switzerland}
\author{Christopher J. Mundy}
\email{chris.mundy@pnnl.gov}


\date{\today}

\begin{abstract}
   First principles molecular dynamics simulation protocol is established using revised 
  functional of Perdew-Burke-Ernzerhof (revPBE) in conjunction with Grimme's third generation 
  of dispersion (D3) correction to describe properties 
  of water at ambient conditions. This study also demonstrates the 
  consistency of the structure of water across both isobaric (NpT) and
  isothermal (NVT) ensembles. 
  Going beyond the standard structural benchmarks for liquid water, we compute
  properties that are connected to both local structure and mass density fluctuations that are related to 
  concepts of solvation and hydrophobicity.  
  We directly compare our revPBE results to  the Becke-Lee-Yang-Parr (BLYP)
  plus Grimme dispersion corrections (D2) and both the empirical 
  fixed charged model (SPC/E) and many body interaction potential model (MB-pol) 
  to further our understanding of how  the computed properties herein 
  depend on the form  of the interaction potential.
\end{abstract}

\pacs{}

\maketitle 

\section{Introduction}
\label{sec:intro}
 Understanding the benefits and limitations of {\emph ab initio} approaches
 based in quantum density functional theory (DFT) for describing 
 aqueous phase  processes in  bulk and in the vicinity of interfaces continues to be an 
 active area of research. Many studies regarding the accuracy of DFT
 to describe both bulk and interfacial properties of neat water
 have been performed and focus on how the role of simulation protocol
 affects  the computable observables.\cite{Yoo_jcp_09,Yoo_jcp_11,Mcgrath_jpca_06,Lin_jctc_12,Akin_cpl_2011,Schmidt_jpcb_09,Ben_jpcl_13,Ma_jcp_12,Jonchiere_jcp_15,Morawietz_pnas_6,Remsing_jpcl_14,Wang_jcp_11,Dion_prl_04,Kessler_jpcb_15, Kuhne_jpcl_10, Baer_jcp_2011,Willard_jpcb_10}   
 More recently, the efficacy of DFT-based methods to describe water in environments ranging 
 from the gas to the condensed phase has been called into question.\cite{Medders_jcp_15, Cisneros_cr_16}
 One solution to the problem is to use  a sophisticated
 classical empirical interaction potential based on a fit to the energetics of
 configurations obtained with high-level wavefunction methods. \cite{Babin_jpcl_12,Medders_jctc_13,Fanourgakis_jcp_08, Burnham_jcp_08}
 These recent studies have produced excellent agreement with structural
 and spectroscopic properties of water and are designed to be correctly
 coupled with path integral calculations to explore the role of nuclear quantum
 effects.
 
 The advantages of using an empirical representation interaction over DFT based methods is clear 
 from the point of efficiency. Until recently the phase behaviour of DFT
 based methods has been informed by relatively short simulation times and
 small system sizes. \cite{Yoo_jcp_09,Yoo_jcp_11,Mcgrath_jpca_06} The results of these studies produced interesting
 results pertaining to the melting points and boiling points of
 popular DFT functionals.\cite{Yoo_jcp_09,Yoo_jcp_11,Mcgrath_jpca_06}  It should be noted that earlier studies of
 these thermodynamic properties were performed with exchange-correlation (XC)
 functionals that did not contain a correction for long-range dispersion 
 interactions that are absent from DFT. \cite{Yoo_jcp_09} 
 
 The recent empirical corrections
 due to Grimme \cite{Grimme_jcc_06,Grimme_jcp_10} have greatly enhanced the agreement with experiment over
 a range of structure, dynamic, and thermodynamic
 properties.\cite{Lin_jctc_12,Yoo_jcp_11,Akin_cpl_2011,Schmidt_jpcb_09,Ben_jpcl_13,Ma_jcp_12,Jonchiere_jcp_15}One of the most important thermodynamic properties
 of DFT water that was markedly improved was the mass density at ambient
 conditions. \cite{ Lin_jctc_12,Schmidt_jpcb_09,Ben_jpcl_13,Ma_jcp_12}This 
 improvement in the mass density using the 
 empirical corrections to the dispersion interaction has allowed for rapid progress to be
 made in the understanding of ions and reactivity in the vicinity of the
 air-water interface.\cite{Baer_jpcb_14}  

 In similar spirit to the fitting empirical potentials to high-level
 wavefunction methods discussed above, empirical interaction potentials 
 using DFT-based levels of electronic structure with and without dispersion
 have been constructed.\cite{Morawietz_pnas_6} These potentials have afforded
the opporunity to perform simulations for relevant
 times-scales and system sizes to converge properties of bulk liquid
 water.\cite{Morawietz_pnas_6} The results of this study further corroborates some past careful
 studies using DFT interaction potentials\cite{Lin_jctc_12,Yoo_jcp_11,Akin_cpl_2011,Schmidt_jpcb_09,Ben_jpcl_13,Ma_jcp_12,Jonchiere_jcp_15} and clears up many inconsistencies
 regarding the thermodynamic properties of DFT water. This aforementined study also suggests 
 a picture where the revised functional  of Perdew-Burke-Ernzerhof (revPBE) \cite{Zhang_prl_98} in 
 conjunction with Grimme's third generation of dispersion (D3) produces
 an effective description of liquid water over a range of condensed phase
 environments.  

 One reason to consider an alternative to parameterized empirical potentials
 is to understand processes that involve the response of liquid water to
  a notional interface. Here, we desire to exploit the flexibility of 
 DFT based interaction potentials to correctly describe the short-range response to 
 an arbitrary perturbation from the bulk liquid, namely solutes, or macroscopic interfaces.  
 To this end, the short-range response to hard sphere 
 cavities of various sizes obtained with DFT was directly compared to
 two popular fixed charged empirical potentials. \cite{Remsing_jpcl_14}
 This study suggests that the quantitative
 differences observed in the short-range response between different water models
 leads to questions about the quality of interaction potentials needed
 to be obtain solvation free energies of ions.  Indeed, an earlier
 study on the local structure of ions as determined by the extended x-ray adsorption
 fine structure (EXAFS) technique found that DFT based interaction potentials
 were required in order to reproduce the accurately measured short-range
 structure.\cite{Bayer_jpcl_11}  Although progress is being made toward
 high-quality empirical force fields for ions based on fits to high-level wavefunction 
 methods,\cite{Arrieta_jpcb_16,Riera_pccp_2016} to the extent that empirical 
 potentials can reproduce the details
 of local solvent response to interfaces remains important 
 research.  
 
 The main focus of previous detailed studies of DFT based methods have been  equilibrium
 structural and dynamical properties.\cite{Morawietz_pnas_6}   Herein, we compare and contrast 
 empirical potentials against DFT for phenomena
 that are germane to computing  solvation free energies, namely mass density
 fluctuations.
 The choice of
 empirical potentials for this study are the SPC/E and MB-pol models of water.
 The former is chosen because of both its popularity and use in the study
 of hydrophobicity; the latter is chosen because of its demonstrated accuracy in producing
 the correct potential energy surfaces as benchmarked by high-level
 wavefunction methods.  This will require
 that we establish the DFT simulation protocol to quantify the role of mass
 density fluctuations 
 under both isothermal (NVT) and isobaric (NpT) ensembles for system sizes
 that are relevant to DFT studies.  The importance of capturing the
 mass density fluctuations at short and long length scales
 forms the corner stone of the theory of hydrophobicity and solvation.  
 Furthermore, the examination of  fluctuations
 provides an additional self-consistent check on the thermodynamic properties of surface tension 
 and isothermal compressibility.\cite{Triezenberg_prl_72,Nilsson_jpl_12}  Going beyond traditional
 probes of aqueous structure,  we contrast the local structure of ambient water 
 by examining the distribution of the 5$^{\rm th}$ nearest neighbor 
 distance ($d_5$). This order parameter was found to be
 relevant for describing the experimental structure of water under pressure 
 and possibly a  diagnostic for providing  signatures of differences between
 empirical  and DFT models of liquid water.\cite{Skinner_jcp_16}
 The goal of this study is to provide a clear comparison of mass density 
 fluctuations between different representations of interaction.  This will
 require the development of  DFT simulation protocol that provides
 a robust and consistent picture of structure and their fluctuations.
 Thus, further advancing  our understanding of 
 the utility of using quantum descriptions of interaction based in DFT 
 to inform our understanding of complex phenomena in the condensed phase.
 \section{Computational details}
\label{sec:comp}
All the simulations presented here have been carried out using the CP2K program within the 
 Born-Oppenheimer approximation \textit{i.e.} the wavefunction was optimized to the ground state 
 at each time step. The QUICKSTEP module within CP2K was used to employ the 
 Gaussian and plane wave (GPW) method.\cite{Lippert_molp_97,Vande_cpc_05} In
 this GPW method both the gaussian and plane wave basis are used to linearly expand molecular 
 orbitals and electronic density, respectively.  
Our model system consisted of 64 water molecules in a cubic simulation box
under periodic boundary conditions. All NpT simulations were carried out at the
ambient thermodynamic conditions, namely the temperature was set to 300 K and the pressure was set to 1 bar
using the reversible algorithm due to Tuckerman and co-workers.\cite{Tuckerman_jp_06} 
The time step was maintained to be 0.5 fs. Nose-Hoover thermostats were employed to all degrees of 
freedom using the ``massive" thermostatting. The time constant of the thermostat and the barostat 
was set to be 11.12 fs (corresponding to 3000 cm$^{-1}$) and 300 fs, respectively. 
All the NVT simulations were carried out using 256 water molecules in a cubic box of side 
length of 19.7319 \AA~  providing a density of 0.997 g/cm$^3$ at a temperature of 300 K.  
Both revPBE\cite{Zhang_prl_98} and BLYP \cite{Becke_pra_88,Lee_prb_88} functional 
were used with the Grimme dispersion correction\cite{Grimme_jcc_06,Grimme_jcp_10} known 
as D3 and D2, respectively. The core electrons were replaced by the norm-conserving pseudopotentials 
of Goedecker and co workers (GTH) \cite{Goedecker_prb_96} to 
carry out the simulations efficiently.  Two types of basis set were used,
a triple-$\zeta$ valence 
Gaussian basis set augmented with two sets of $d$-type or  $p$-type polarization functions (TZV2P) and the 
molecularly optimized double-$\zeta$ basis set (MOLOPT-DZVP-SR-GTH which we
will refer to as MOLOPT in the remaining text)\cite{Vande_jcp_07}.  
Both of these basis sets were previously successfully used with these functionals in the NVT simulations
of bulk water at ambient, high pressure, and high temperature conditions.\cite{Bankura_jpcc_14,Skinner_jcp_16} 
In a NpT molecular dynamics run, longer simulation times are required to obtain the equilibration and to 
sample the fluctuation. We ran the simulations to produce a 100 ps long trajectory, from which the last 50 ps 
was used to gather statistics.
The NpT dynamics were carried out using a larger reference simulation cell to
ensure a constant number of grid points and provide a lower bound on
the electron density cutoff. The reference simulation cell used was 19 \% larger 
than the original simulation cell, and corresponding to the density of 0.59 g/cm$^3$. 

\subsection{Establishing the NpT simulation protocol}

It has been established that the cutoff of 400 Ry for the
expansion of electron density in the planewave basis produces converged
results in the NVT ensemble. However, in the case of 
NpT ensemble, a much larger cutoff is needed to produce the converged virial. 
It has been reported by McGrath et al. \cite{Mcgrath_cpc_05} that an NpT  Monte Carlo simulation 
with a cutoff of 1200 Ry produced 10\% lower density than that with a cutoff of
280 Ry. Another more recent NpT  Monte Carlo simulation by Del Ben \textit{et. al.} used a cutoff 
of 800 Ry.\cite{Ben_jpcl_13} They confirmed that changing cutoff from 800 to 1200 Ry did not affect the density. 
In the original NpT MD simulation, Schmidt \textit{et. al.} found that increasing cutoff from 
600 to 1200 Ry did not change the density.\cite{Schmidt_jpcb_09} 

However, there are many options for simulation in the NpT ensemble within CP2K
and it is instructive to provide useful information regarding how simulation
protocol can effect the outcome.  A summary of these options in addition to 
convergence tests are detailed in Appendix A.
By using the standard Fourier interpolation technique 
the total pressure (as defined by $\frac{1}{3}{\rm Tr}{\bf \Pi}$, where
${\bf \Pi}$ is defined in Ref.\citenum{Schmidt_jpcb_09})
was sufficiently converged to at a cutoff of 800 Ry to reproduce a mass 
density in agreement with previous studies (see Appendix A).
\begin{figure}
\includegraphics{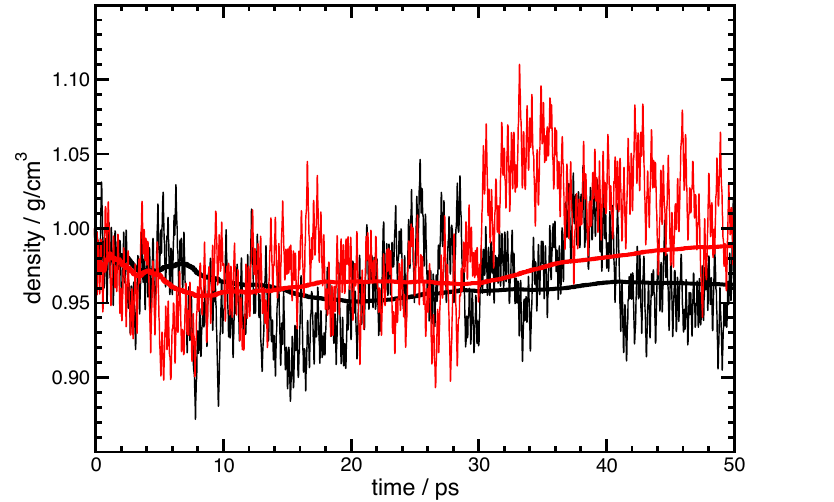}
  \caption{\label{fig:3}The instantaneous density fluctuation and its running average as a function of
 simulation time from the NpT simulation at revPBE-D3/TZV2P (black) and revPBE-D3/MOLOPT (red) level of theory. }
\end{figure}

\section{Results and discussion}
\label{sec:res}
\subsection{Structural distributions}
\label{subs:small}
\subsubsection{Mass density and radial distribution functions}
Having established the simulation protocol used in this study, we can turn to
the calculation of the mass density of DFT based interaction potentials.   
The mass density  can be calculated from an NpT run, using the aforementioned
protocol, by taking an average of the 
instantaneous fluctuating volume over the simulation time.  Figure \ref{fig:3} shows the variation of 
instantaneous mass density and the corresponding running average 
with simulation time for the 64 water box using revPBE-D3 functionals with TZV2P and MOLOPT basis set. 
The calculated average value and the root mean square deviation are given in the Table \ref{tab:1} .  
Our estimates provide a picture  where the revPBE-D3 functional is providing
a density of 0.962 g/cm$^3$ and 0.988 g/cm$^3$ with TZV2P and MOLOPT basis set, respectively. These values are 
in good agreement 
with the experimental density of 0.997 g/cm$^3$. The difference of 0.01-0.03
g/cm$^3$ does not account 
for more than 1 \% in the
lattice constant making up the simulation supercell. 

\begin{table*}
\caption{Density, compressibility and structural data obtained from NpT simulations at various level of 
  theories for bulk water at ambient conditions}
\begin{tabular}{lllll}
\hline
  Property& revPBE-D3/TZV2P& revPBE-D3/MOLOPT& BLYP-D2/TZV2P& Exp\\
  \hline
  $\rho$ (g/cm$^3$) & 0.962$\pm$0.029& 0.988$\pm$0.040 &1.04$\pm$0.026 &0.997\\ 

   $\kappa_T$ (Mbar$^{-1}$) &  42  &  -- & 35 &45 \\ 

 1st max  r [\AA~] &  2.80  & 2.82  & 2.75 &2.80 \\ 

 1st max g$_{OO}$(r)  & 2.74   & 2.50  & 3.24 & 2.57 \\ 

 1st min  r [\AA~] &  3.45  & 3.66  & 3.35 &3.45 \\ 

 1st min g$_{OO}$(r)  &  0.82  & 0.91  & 0.72 &0.84 \\ 
\hline
\end{tabular}
\label{tab:1}
\end{table*}%

Our calculated value of 0.96 g/cm$^3$ is consistent with the previously reported value of 
0.96  for revPBE-D2 by Lin et al.\cite{Lin_jctc_12} However, in the cited study
they did not calculate the density directly 
from an NpT ensemble. Instead, they used an indirect method where the total
energy was calculated 
as a function of the scaled lattice constance for a given snapshot obtained
with a Car-Parrinello molecular dynamics (CPMD) trajectory. The equilibrium 
mass density was obtained from the minimum of the interpolated energy. 
Our values are less than the previously
reported value of 1.02 g/cm$^3$ by Wang et al. \cite{Wang_jcp_11} for revPBE using
the nonlocal van der Waals (vdW) correlation functional 
proposed by Dion et al\cite{Dion_prl_04}.  However, these calculations are also
not obtained from a traditional NpT
ensemble. Rather, this study calculated the equilibrium density from the
pressure-density curve obtained from NVT simulations at 
different volumes. To the best of our knowledge, our results report 
the first NpT simulations
of revPBE-D3 water and its equilibrium density at ambient conditions. 
Like other popular gradient corrected (GGA) functionals (e.g. PBE and BLYP), in 
the case of revPBE-D3 the density has been significantly improved (from 0.69 to 0.96) towards the experimental value 
with the inclusion of dispersion correction (Grimme D3). This is consistent with the 
consensus that GGA functionals require the dispersion corrections to obtain
a physically reasonable description of liquid water.

\begin{figure}
\includegraphics{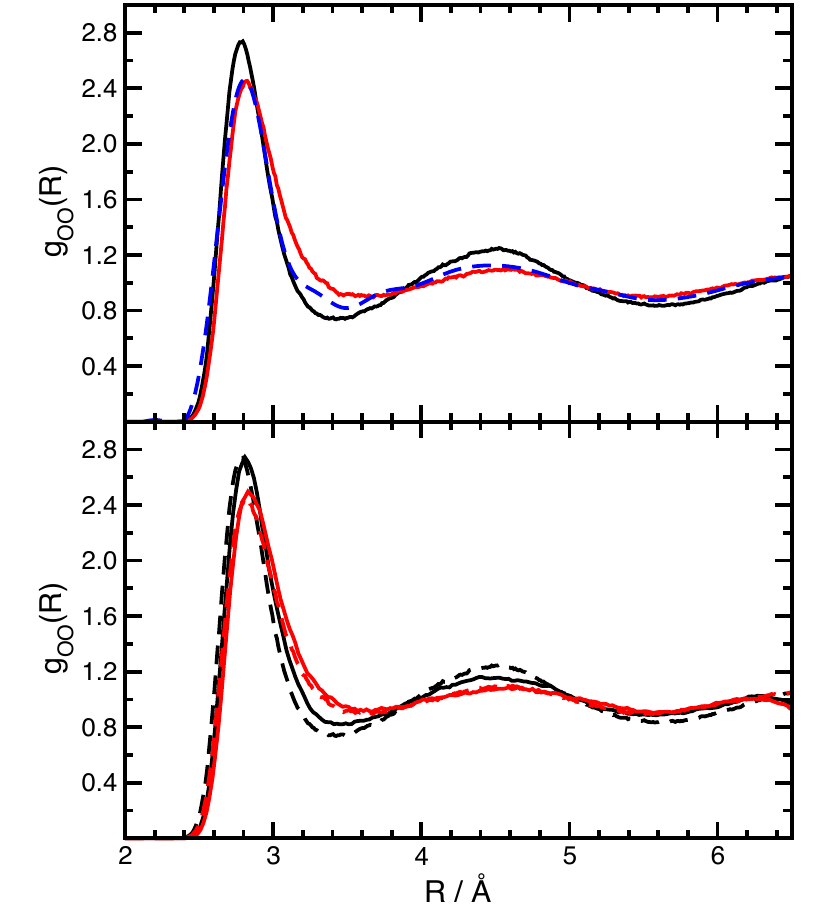}
  \caption{\label{fig:4}RDFs  for oxygen-oxygen distances: a) RDFs obtained from NpT 
simulations at revPBE-D3/TZV2P (black) and revPBE-D3/MOLOPT (red) basis sets, compared to the 
experimental RDF (blue dashed) obtained from XRD  (taken from Ref. \citenum{Skinner_jcp_13}) and b) RDFs obtained 
from NpT (solid lines) and NVT (dashed lines) ensembles for revPBE-D3/TZV2P (black) and 
revPBE-D3/MOLOPT (red) level of theory. }
\end{figure}

Figure \ref{fig:4} top panel depicts the oxygen-oxygen radial distribution
functions (RDF) from our NpT
simulations using both TZV2P and MOLOPT basis sets along with the experimental radial distribution functions 
extracted from Ref. \citenum{Skinner_jcp_13}. Our calculated 
RDF using the TZV2P basis set shows an excellent
agreement with the experimental data. 
Most importantly, the position of the first peak is in the
correct position, and the first minimum contains the correct amount of
disorder as compared to experiment. This suggests that revPBE-D3 water
has the potential to display  better diffusivity at 300K as compared to other popular
GGA functionals. Previous Monte Carlo simulations in the NpT ensemble for BLYP-D3, PBE0-ADMM-D3, and MP2 
have predicted the first minimum in the RDF to be significantly more shallow
than the experiment although a good mass density is reproduced.\cite{Ben_jpcl_13} 
The direct calculation of the diffusion constant is beyond the
scope of this paper as it would require many long trajectories in the NVE
ensemble. The only significant deviation from the results herein is the height of the first peak that
is higher by 0.2  when  compared to experiment. It should be noted that we did not
include the nuclear quantum effect (NQE) into our simulations. It was previously reported that 
inclusion of NQE might influence the height of the first peak towards the experimental 
value.\cite{Kuharski_jcp_85,Lobaugh_jcp_97,Mahoney_jcp_01,Morrone_prl_08,Ceriotti_cr_16}

RDFs calculated using the short-range molecular optimized basis set
(MOLOPT)\cite{Vande_jcp_07} at the double-$\zeta$
level is also in good agreement with the experiment. Interestingly, MOLOPT 
produces the correct height of the first peak but is slightly shifted to larger
distances than the experiment.
Moreover, the first minimum suggests less structuring as compared 
to both TZV2P and the experimental results. To understand the origins of the
difference between the two basis sets, we compared the RDF calculated with and without the D3
dispersion correction (see the supporting information for the corresponding
RDFs). Our calculations indicate that in the
absence of the dispersion correction both the basis sets give similar RDFs, however in the presence of the D3 dispersion
correction, MOLOPT results deviate from that of TZV2P. This indicates that the origin of the difference is due to 
the matching of the basis set with the Grimme dispersion correction scheme. Since, the original D3 parameters
were optimized with TZV2P basis sets, they don't work as well with the MOLOPT basis sets.

As another self-consistent check of our NpT protocol, we compare the RDF calculated from our 
NpT ensembles to those calculated
from NVT ensembles. Theoretically, 
the NVT and NpT approach should yield the same results if the protocol  in both
approaches is converged.
Indeed, our RDFs from both NpT and NVT simulations are similar as shown in the bottom panel of Fig \ref{fig:4}.  This is 
a clear improvement in our understanding  between the different approaches to
simulation. 
Previous results for MP2 water using NpT Monte Carlo simulations provided a very different RDF 
than that obtained by  simulations using other ensembles.\cite{Ben_jpcl_13,Vande_jcp_05}

\begin{figure}
\includegraphics{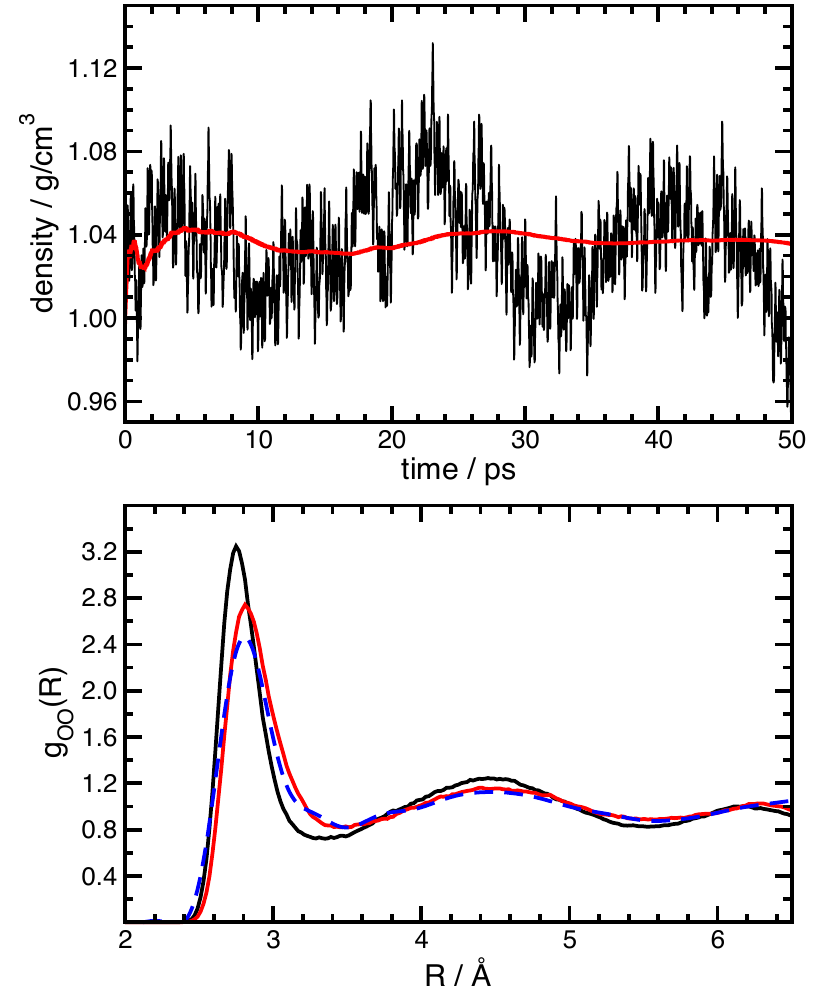}
  \caption{\label{fig:5}
  a) The instantaneous density fluctuation and its running average as
  a function of simulation time from the NpT simulation at the BLYP-D2/TZV2P level
  of theory; b) RDFs for oxygen-oxygen distances at the BLYP-D2/TZV2P (black)
  level of theory, compared to that at revPBE-D3/TZV2P (red) level of theory and
  the experimental RDF (blue dashed) obtained from XRD (taken from Ref.
  \citenum{Skinner_jcp_13}).
  }
\end{figure}

Additional comparisons between revPBE-D3 and BLYP-D2 were carried out in the
NpT ensemble. BLYP-D2 has been a popular choice for numerous past studies of water 
and is known to produce satisfactory results regarding the mass
density.\cite{Schmidt_jpcb_09,Bankura_jpcc_14,Mcgrath_pccp_11} Our simulations suggest that the mass density obtained using
BLYP-D2 at 300K is
1.04 g/cm$^3$ (see Figure \ref{fig:5} top panel). This is slightly higher than the reported value of
0.992 g/cm$^3$ ($\pm$0.036) by Schmidt et al.\cite{Schmidt_jpcb_09} This deviation might be attributed to the 
difference in the temperature (330K used by Schmidt \textit{et. al.}). 
The slightly higher density obtained here with  BLYP-D2  (1.04 g/cm$^3$)  is also comparable to
the BLYP-D3 density reported by Del Ben \textit{et al.} (1.066 g/cm$^3$)
\cite{Ben_jpcl_13} and by Ma \textit{et al.} (1.07 g/cm$^3$) \cite{Ma_jcp_12}. 

The bottom panel of Figure \ref{fig:5}  compares the oxygen-oxygen RDFs for revPBE-D3 and 
BLYP-D2, both using TZV2P basis sets, along with the experimental RDF. The height of the 
first peak for BLYP-D2 is too pronounced 
and the first minimum is also significantly deeper when compared to the
the experimental results. Moreover, the position of the first peak is also slightly 
at a lower distance compared to the experiment. Overall, our research suggests
that revPBE-D3 is producing a better overall mass density and liquid structure as determined by
the experimental oxygen-oxygen RDF than BLYP-D2.

\subsubsection{Local structure}
\label{subs:convsstep}
The results of the previous section suggest there is no significant
distinction between the revPBE-D3  and the empirical
parameterizations of water considered in this study.  Going beyond the RDF to
provide a more detailed description of the local structure of water is one way
to differentiate between different representations of interaction for water.  
Understanding the local structure is crucial to understand the anomalous 
thermodynamic and kinetic behavior of water. Elucidating whether the 
local structure of water is just a random collection of states generated by hydrogen 
bond fluctuations or the competition between different specific 
locally favored structures in the free energy landscape could play a key role
in advancing our understanding of the bulk homogeneous phase of
water.  To characterize the local structure of water, a wide range of different order
parameters have been used.\cite{Dijon_jpcb_15} The most widely used order parameter is the 
tetrahedral order parameter ($q$) that is focused only on the first shell water. 
Previous studies have shown that order parameters that describe the second shell 
order may play a role in our understanding of the bulk homogeneous phase of water.\cite{Dijon_jpcb_15} 
In a recent work,\cite{Skinner_jcp_16} it was observed that the behaviour of 
the 5$^{\rm th}$ nearest neighbour water molecule is crucial to understand the change in local 
structure of water under pressure (from ambient to 360 MPa pressure). The
so-called $d_5$ order parameter has been previously used to investigate the local 
structure of supercooled water.\cite{Cuthbertson_jcp_11,Singh_jcp_16} 
Here, we have focused on the distance of the 5$^{\rm th}$ water from the central 
water molecule as a suitable order parameter to analyze the local structure of 
water at ambient conditions. 

\begin{figure}
\includegraphics{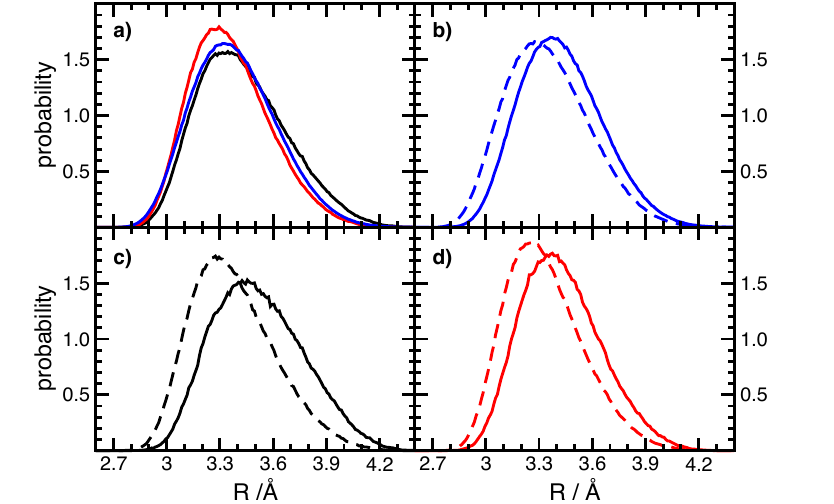}
\caption{\label{fig:8}
Probability distribution of a) d$_5$ order parameter in bulk ambient water obtained from the NVT simulations at the revPBE-D3/TZV2P (black), MB-pol (red) and SPC/E (blue) level of theory. Distribution of d$_5$ based on when the 5th water is hydrogen bonded (solid line) to any of the first shell waters vs. that when it is not hydrogen bonded (dashed line) for b) SPC/E, c) revPBE-D3 and d) MB-pol. 
  }
\end{figure}

To this end, we have analyzed the revPBE-D3, SPC/E, and MB-pol (in the NVT ensemble
under bulk periodic boundary conditions in a supercell containing
256 water molecules) in terms of this $d_5$ order parameter. 
We have calculated $d_5$ as follows: For a water molecule $i$, 
we ordered all the other water molecules in the 
simulation according to the increasing radial distance to that water oxygen
from the $i$-th water oxygen ($d_{ji}$).
Then the order parameter $d_5$ is simply the distance between the $i$-th water
oxygen and its 5$^{\rm th}$ water oxygen ($d_{5i}$).
The first panel in Figure \ref{fig:8} displays the probability distribution of $d_5$ over all the
water molecules in the simulation box for the revPBE-D3, MB-pol and SPC/E
waters. The average value of $d_5$  for the revPBE-D3, MB-pol and SPC/E water at
ambient conditions are 3.49 \AA, 3.36 \AA~ and 3.38 \AA~, respectively.  
Although all models show quantitative differences, SPC/E and revPBE-D3 are in better agreement than
with MB-pol.

We can take this analysis a step further and consider the influence of the 
hydrogen bonding between the 5$^{\rm th}$ water
molecule to any of the four water molecules comprising first solvation shell.
Specifically, we calculated the $d_5$ for all water molecules in the
simulation and divided them into two groups. The first group  represents
configurations in which the 5$^{\rm th}$ water forms a hydrogen bond to any of the first
shell water.  The second group represent the 5$^{\rm th}$ that is not hydrogen bonded to molecules 
comprising the first shell.  We use the 
standard hydrogen bond criteria of the distance between two
oxygens being less than 3.5 \AA~ and O-H-O angle being less than
30$^{\circ}$. 
The results are shown in the remaining three panels of 
Figure \ref{fig:8} where the classical empirical
potentials show distinct behavior when compared to revPBE-D3.
However, when examining the hydrogen bonding distributions of the 5$^{\rm th}$
water, MB-pol and revPBE-D3 seem to be in better qualitative agreement.
The computed distances between the 5$^{\rm th}$ water that is 
hydrogen bonded remains further away from its' partner oxygen
with the average value of 3.51, 3.43 and 3.40 \AA~, for DFT, MB-pol, and
SPC/E water, respectively.  Non-hydrogen bonded distances are 
3.46, 3.33 and 3.36 \AA~, for the DFT, MB-pol and SPC/E water, respectively. 
Although all absolute distances are different, there seems to be the largest
effect between hydrogen bonding and non-hydrogen bonding in the MB-pol
representation of interaction.
\begin{figure}
\includegraphics{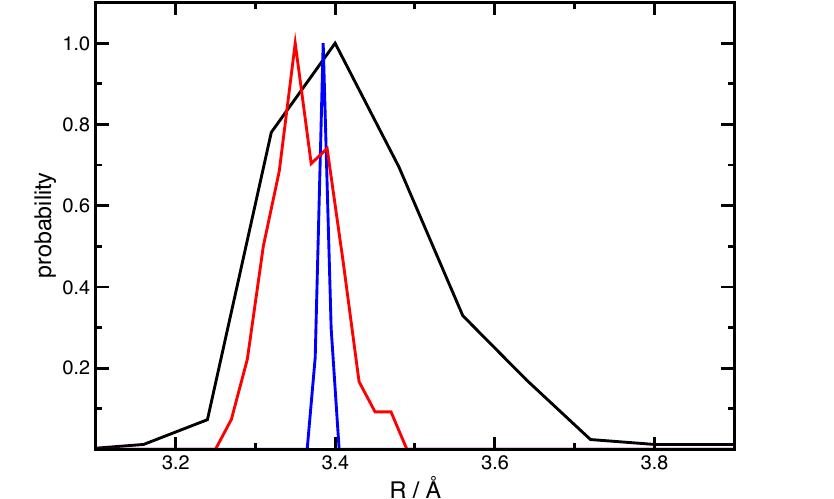}
\caption{\label{fig:9}
Probability distribution of the average value of d$_5$ order parameter for each individual water in the simulation box  obtained from NVT simulations of revPBE-D3/TZV2P (black), MB-pol (red) and SPC/E (blue) level of theory.
  }
\end{figure}

Finally, we can look at the average value of $d_5$ for each individual water
molecule. 
Figure \ref{fig:9} displays
the distribution of the mean value of $d_5$ for each individual water for all three models. 
The salient point of this analysis is that it clearly demonstrates the
inflexibility of the SPC/E water model predicting  a very narrow
distribution of the average $d_5$.  This is an indication that every SPC/E
water molecule has nearly the identical average environment.
On the other hand,  DFT water produces a wide distribution in $d_5$ 
suggesting that on average water explores a wide range of local environments  even under
bulk homogeneous conditions. 
The MB-pol model seems to capture this local heterogeneity and has a broader distribution than SPC/E 
but remains significantly more restricted than DFT.
Capturing the flexibility of  $d_5$ under ambient conditions seems to be an indication of
the ability of a water model to describe the correct structure under different
environments\cite{Skinner_jcp_16}.  To the extent that this is a relevant
distribution under bulk homogeneous conditions at ambient  conditions is yet to
be determined experimentally.  Nevertheless, this order parameter 
that is presented here  is able to discern between different
representations of interaction.  
\subsection{Mass density fluctuations}

\subsubsection{Isothermal compressibility}
\label{subs:big}
\begin{figure}
\includegraphics{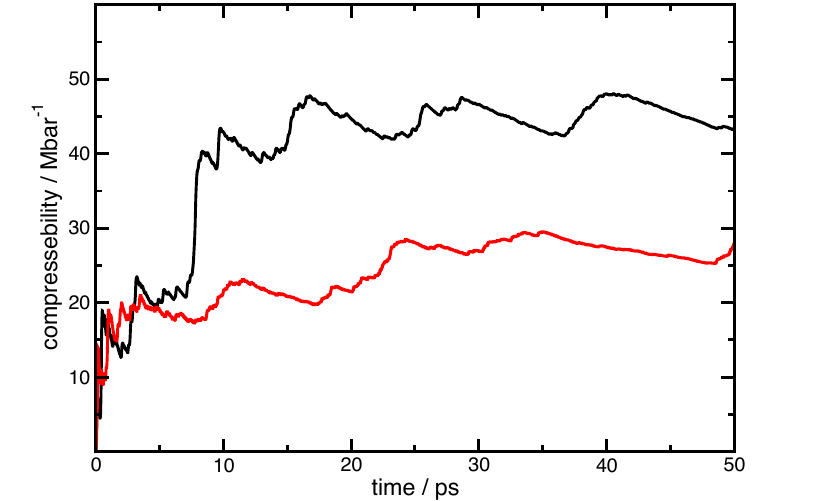}
\caption{\label{fig:6}
The running average of the isothermal compressibility calculated from the NpT simulation  at the 
revPBE-D3/TZV2P (black) and BLYP-D2/TZV2P (red) level of theory.
}
\end{figure}

Now that the properties of two DFT based interaction potentials have been established 
with respect to experimental radial distribution functions and mass density, we
can further push our understanding of water
examine and compare the quality of the mass density fluctuations
that are related to thermodynamic variables.
To start, we examine the isothermal compressibility, $\kappa_T$  using the instantaneous
volume fluctuations in the NpT ensemble using the following formula:
\begin{equation}
  \kappa_T=\frac{\left<V^2\right> -\left<V\right>^2}{k_{\rm B}T\left<V\right>}
\end{equation}
 where $V$ is the instantaneous volume of the system, $k_{\rm B}$ is Boltzmann
 constant and $T$ is the simulation temperature.
The volume fluctuations were averaged over a trajectory. Figure \ref{fig:6} 
depicts the variation of the isothermal compressibility as a function of simulation time. 
An average over the 50 ps trajectory provided a value of 42 Mbar$^{-1}$ for revPBE-D3 and 
35 Mbar$^{-1}$ for BLYP-D2. The revPBE-D3 value, as expected, is very close to the experimental value of 45 Mbar$^{-1}$
and is in much better agreement 
than the previously computed value using PBE-D3 (21 Mbar$^{-1}$) and for PBE0-D3 (32 Mbar$^{-1}$) 
by Gaiduk \textit{et al.}\cite{Gaiduk_jpcl_15} 
Interestingly, both PBE-D3 and PBE0-D3 were reported to produce a density of 
water very close to the experimental density (1.02 g/cm$^3$ and 0.96 g/cm$^3$, respectively). 
revPBE-D3 water also produces a compressibility 
in better agreement with experiment  than the vdW-DF, vdW-DF$^{\rm {PBE}}$ and VV10 level of 
theory (18.2, 32,2 and 59.0 Mbar$^{-1}$, 
respectively) as computed by Corsetti \textit{et al.}\cite{Corsetti_jcp_04} 
It is  interesting to note  that among the four popular classical empirical
water models  (\textit{i.e.} SPC/E, TIP4P, TIP4P/2005, and TIP5P), SPC/E and
TIP4P/2005 is found by Helena \textit{et al.}
to produce the best agreement with the experimental compressibility at ambient
conditions.\cite{Helena_mp_09} 
However, SPC/E fails  to produce the change of compressibility with temperature.
TIP4P/2005 is the only point charge model parameterized to produce the
temperature dependence of the isothermal compressibility
for liquid water.\cite{Helena_mp_09}
The MB-pol model produced an isothermal compressibility of 45.9 Mbar$^{-1}$ at ambient conditions which is in excellent agreement with experiment. \cite{Reddy_jcp_16}
MB-pol model was also found to produce the experimental pattern of changing isothermal compressibility with temperature.\cite{Reddy_jcp_16} 
As an additional self-consistent check to the estimates of the isothermal
compressibility that were provided above, we examine the impact of both the thermostat and barostat 
frequency on our results. 

\begin{figure}
\includegraphics{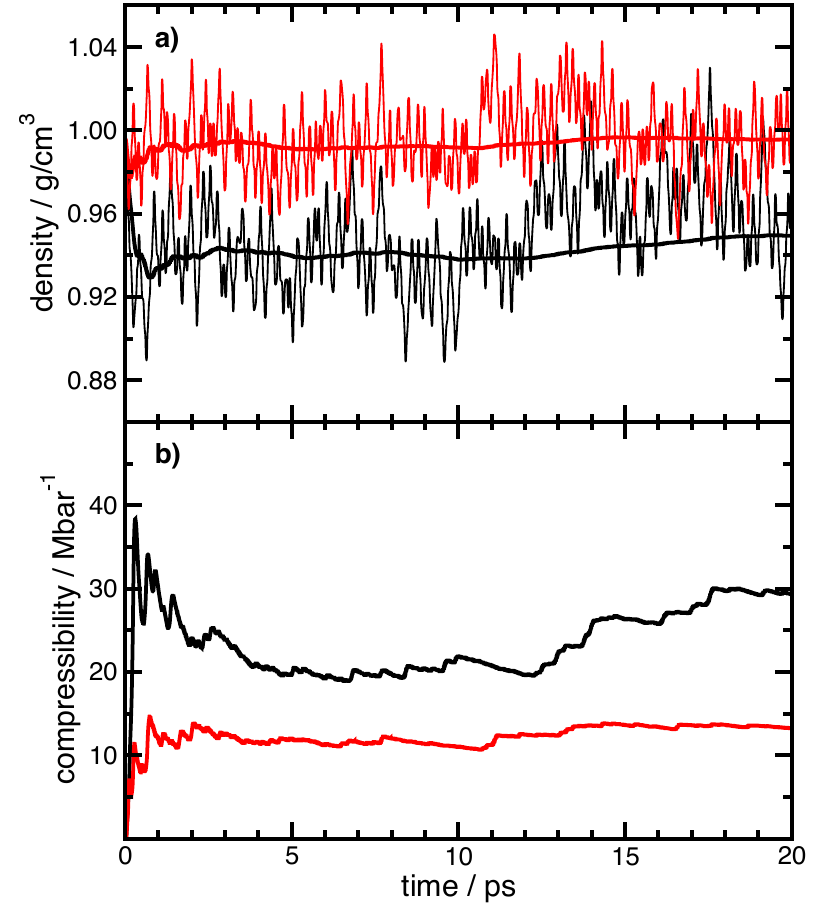}
\caption{\label{fig:7}
a) The instantaneous density fluctuation and its running average 
as a function of simulation time  and b) the isothermal compressibility as a function of simulation time. 
Both were calculated from the NpT simulation at the revPBE-D3/TZV2P (black) and BLYP-D2/TZV2P (red) level 
of theory with thermostat and barostat frequency of 1 ps.
  }
\end{figure}

To this end, we have conducted separate simulations of 20 ps in length using a significantly
higher time constant (\textit{i.e.} lower frequency) for both namely, 1 ps. Figure \ref{fig:7} shows the computed
instantaneous mass density 
and the isothermal compressibility for these revised simulations.  As shown in
Figure~\ref{fig:7}, the revPBE-D3 density 
is not significantly effected by the lower thermostat and barostat frequency.
Interestingly, the BLYP-D2 mass density  
decreased from $\sim$ 1.02 g/cm$^3$ (see Figure \ref{fig:5} top panel) to
$\sim$ 0.99 g/cm$^3$.  Moreover, there seems to be a nontrivial but small
dependence on the choice of thermostat frequency on the resulting mass density
of DFT-based water.  It should be pointed out that the reason for the 
higher barostat frequency was to be able to sample volume fluctuation over the
significantly shorter simulation time afforded by DFT. \cite{Schmidt_jpcb_09}   A more
reasonable barostat time-constant of 1 ps is generally used in conjunction with classical
empirical potentials.  
Not surprisingly,  the 
isothermal compressibility displays a significant dependence on the barostat and
thermostat values  producing lower values for both functionals studied
herein (see Figure \ref{fig:6} and Figure \ref{fig:7} bottom panel).
To further our understanding we examined the isothermal compressibility of the SPC/E water using 
a barostat
both a time-constant of 300 fs and 2 ps. The values for $\kappa_T$ were 44.7  and 45.2
Mbar$^{-1}$ for the lower and higher frequency, respectively. This is not
surprising since SPC/E model is rigid and likely has no significant coupling to either the
low or high frequency barostat. Nevertheless, a more systematic study
on the effects of the coupling between the representation of interaction and the barostat and
thermostat frequencies are needed.

\subsection{Response to the air-water interface}

Having explored the dependence of molecular interaction on  mass density
fluctuations in bulk, we now move to the vicinity of the air-water interface.
It is the mass density fluctuations in the vicinity of the air-water
that provides the direct connection to the surface tension.  Only recently,
has the surface tension of water been computed using an energy based
methodology.  The advantage of examining the fluctuations directly is that we
are not concerned with the convergence of the pressure as was discussed above.
We begin by computing the mean density profile as a function of the distance from the instantaneous 
interface. The air-water interface was simulated using a 20\AA~x 20\AA~x 50\AA~slab.
Three different water models were used, DFT based ab initio model 
(revPBE-D3), empirical many body potential 
model (MB-pol) and classical potential model (SPC/E).
The instantaneous interface 
was calculated using the method proposed by Willard and Chandler and has been
shown to be a superior coordinate for studying interfaces over the widely used 
Gibbs dividing surface.\cite{Willard_jpcb_10} 
Following Willard and Chandler, a coarse-grained 
time dependent density field was defined as: 

\begin{equation}
  \rho({\bf r},t)=\sum_{i=1}^{N} (2\pi \zeta^2)^{-3/2} 
  \exp-{\frac{1}{2}\left(\frac{|{\bf r}-{\bf r}_{i}(t)|)}{\zeta}\right)^2}
\end{equation}
where $\zeta$ is the coarse graining length and $\bf{r}_{i}$ is the position of
$i^{\rm th}$ particle at time 
$t$. Considering the molecular correlation length of water, the 
value of $\zeta$ usually chosen to be 2.4 \AA~. 
The instantaneous surface is defined by the isosurface $h(x,y)$ having a density
equal to half of the bulk density. 
Once the interface is
identified, we then calculate the distance of each water molecule from the instantaneous interface 
for each configuration ($a_{i}$) is as follows:
\begin{equation}
  a_{i}=\{[{\bf s}_{i}(t)-{\bf r}_{i}(t)] \cdot {\bf n}(t)\}
\end{equation}
where ${\bf s}_{i}(t)$ is the $h(x,y)$ for the corresponding ${\bf r}(x,y,z)$
for $i^{\rm th}$ configuration, 
and ${\bf n}(t)$ is the surface normal vector at the $h(x,y)$. 
The mean mass density profile($\rho$) is then calculated as a function of the distance from the instantaneous 
interface($z$) using the following:
\begin{equation}
\rho(z)=\frac{1}{L^2}\langle \sum_{i=1}^{N} \delta (a_{i}-z)\rangle
\end{equation}
where $L$ is the length of the simulation cell and $\delta$ is the Dirac's delta function.

\begin{figure}
\includegraphics{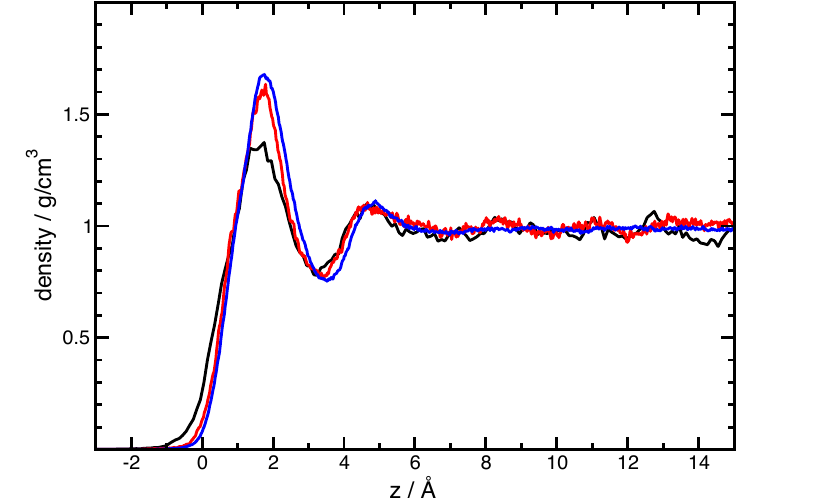}
\caption{\label{fig:10}
Mean density profile as a function of distance from the instantaneous interface
  for revPBE-D3 (black), MB-pol (red) and SPC/E (blue) slab geometries. UNIT needs
  to be correct for density. 
  }
\end{figure}

Figure \ref{fig:10} depicts the mean density as a function of distance from the instantaneous interface
for the water models studied here. 
In general, the density profiles for all three models (i.e. DFT, SPC/E, and MB-pol) show the well-defined
peak with clear minima at the interfacial region indicating that water
molecules are more structured in the vicinity of the interface.
Our work is consistent with previous studies using both classical
and \textit{ab initio} potential.\cite{Willard_jpcb_10,Kessler_jpcb_15} Among the three models 
revPBE-D3 produces a less structured water in the vicinity of the interface than 
either MB-pol and SPC/E that are both in near quantitative agreement.
The results in Figure \ref{fig:10} provide an important self-consistent check on the mass density presented
in Section~\ref{subs:small}. A good estimate of the mass density can be gleaned
from  Figure \ref{fig:10} as it converges to the value of 1 g/cm$^3$  for 
all interaction potentials in this study.  BLYP-D2  simulations that are not
shown here have been performed and have shown a similar excellent agreement
with the value of 1 g/cm$^3$.\cite{Kessler_jpcb_15, Kuhne_jpcl_10, Baer_jcp_2011}  
Our results suggest that all protocols, namely NpT and NVT in slab geometry,
will converge as longer NpT simulations can be performed using barostat
time-constants on the orders of picoseconds.

\begin{figure}
\includegraphics{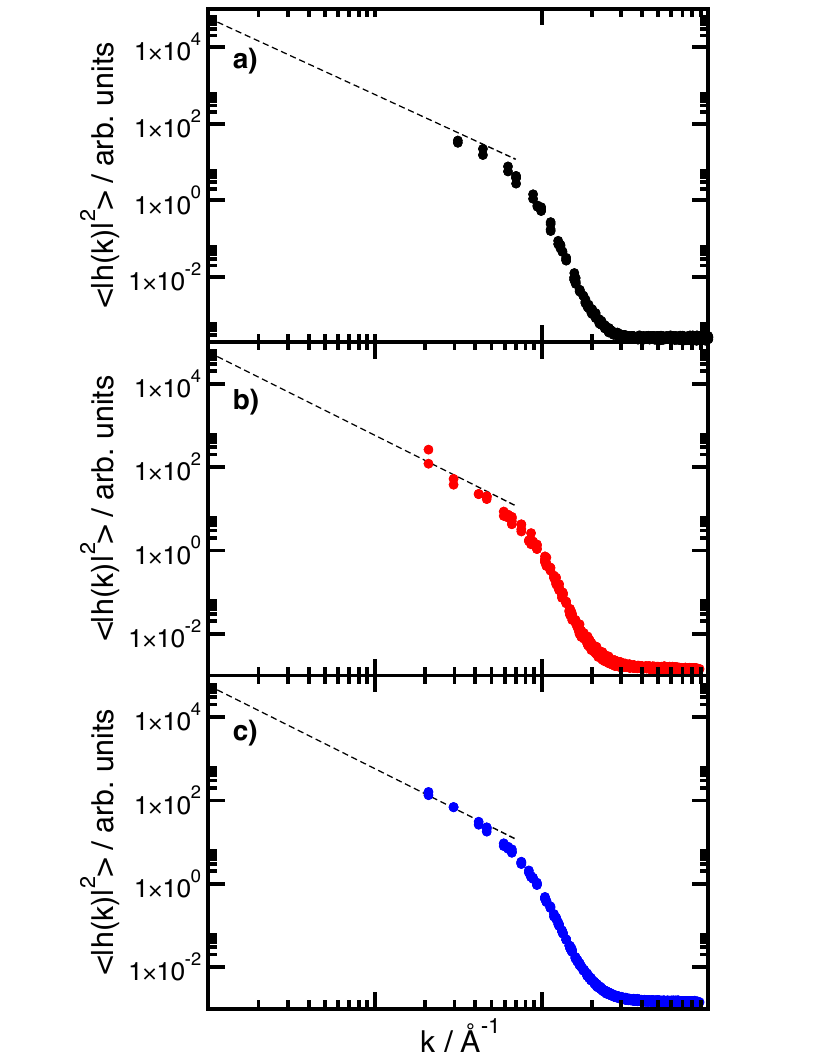}
\caption{\label{fig:11}
Fourier transform of the instantaneous interface configuration h(x,y), ($ \tilde {h} $ (k))
 for a) revPBE-D3, b) MB-pol and c) SPC/E.  The dashed line is the fit for $\gamma=72.0~ mJ/m^2$ to 
the capillary wave theory ($ \left<\right.$$\tilde {h}$ (k)$\left.\right> $$^2$ $\approx$ 1/ $\beta$ $\gamma$ k$^2$ ).
}
\end{figure}

Given that there are some quantitative differences in the mass density profile shown in Figure \ref{fig:10},
it would be useful to provide some measure to how the structural averages,
effect the thermodynamic property of surface tension through an analysis of the
fluctuations in height of the instantaneous interface, namely $h(x,y)$ . 
Figure \ref{fig:11} displays  the power spectrum of the instantaneous water-vapor interface. The
Fourier transform [$\tilde{h}(k)$] of the instantaneous interface $h(x,y)$ and is 
related to the surface
tension through macroscopic capillary-wave theory for wave vectors less than $\approx$2$\pi$/9 \AA~. 
As one can glean from Figure \ref{fig:11}, all models studied herein are nearly
indistinguishable in terms of their height fluctuations even though revPBE-D3 was
produced both a slightly wider and understructured interface as was shown in
Figure \ref{fig:10}.  As a guide, we have plotted the linear response 
curve that is consistent with the experimental surface tension, $\gamma$  of
72.0  mJ/m$^2$ in the range 
0.01\AA$^{-1}$~$< k <$ 0.7 \AA$^{-1}$.
One clearly sees the deviations from linear response as would be expected for
short distances (large $k$) and suggests a picture where all of the models
presented herein provide satisfactory agreement with the experimental surface
tension of water.\cite{Patel_pnas_11} 
In order to be quantitative, one would have to simulate 
much larger surfaces in order to have a significant linear region to extract
a precise surface tension.  It should be noted that the surface tensions of
SPC/E and MB-pol are  63.6\cite{Vega_jcp_07} and 66.8\cite{Reddy_jcp_16} $mJm^{-2}$, respectively.  
However, we are encouraged by 
our results presented in this study as they suggest that the fluctuations at scales 
relevant to modern DFT simulations are accurately represented.

\subsection{Response to microscopic interfaces}

\begin{figure}
\includegraphics{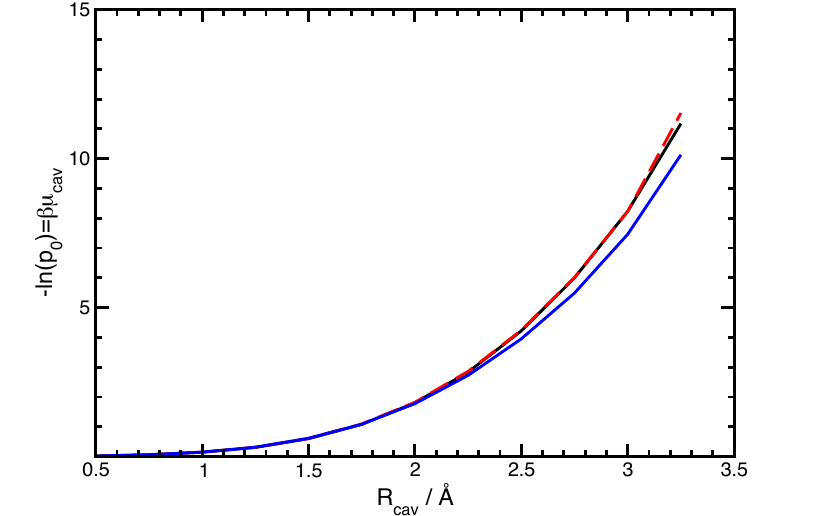}
\caption{\label{cavEnfullfrfig} Cavity formation free energy for three water models as 
  a function of cavity size calculated using the slab geometry revPBE-D3 (MOLOPT) (black solid line), MB-pol (red dashed line) and SPC/E (blue solid line).
  }
\end{figure}

The aforementioned results on surface tension
are probing the response of water models to a large hydrophobic
interfaces at length scales where capillary waves dominate.
The opposite limit of small interfaces is equally important and getting
the balance between small and large length scale response correct forms the
basis of describing the hydrophobic effect. \cite{Chandler2005}
The free energy of forming a large macroscopic interface is given by the surface tension multiplied 
by the surface area of the interface. This relationship breaks down for very small interfaces at the
molecular scale.\cite{Chandler2005} Beyond the hydrophobic effect,  these molecular scale 
interfaces are important for estimating
solvation free energies of small molecules and for the hydrophobic interaction between small molecules
in solution. For small cavities the free energy of cavity formation energy can be estimated with 
the Widom particle insertion formula:
\begin{equation}
\Delta \mu_{X}^{\rm cav}=-k_{\rm B}T\ln \left<\exp^{-\beta U_{\rm cav}}\right>_{0}
\end{equation}
where  $U_{\rm cav}$ is a hard sphere repulsion that acts only on the oxygen atoms out to a
radius of $R_{\rm cav}$, and $\beta=1/k_{\rm B}T$. This expression can be rewritten in terms of the probability of observing
a cavity of a given size in pure water.
 \begin{equation}
\Delta \mu_{X}^{\rm cav}=-k_{\rm B}T\ln p_0(R_{\rm cav})
\end{equation}
We can estimate this energy by monitoring the probability of observing a cavity of a given 
size in a pure water simulation.  Figure~\ref{cavEnfullfrfig} compares this quantity for the
revPBE-D3, MB-pol and SPC/E models. It is calculated using the slab simulation described above.  We see
that MB-pol and revPBE-D3 with the MOLOPT basis set agree over the whole size range studied. SPC/E has
a comparatively lower cavity formation energy for the larger sizes. 
A comparison of the cavity formation free energy  between different DFT
functionals and protocols is discussed
in Appendix B.  It suffices to say that differences between basis sets
and functionals appear at the larger cavity radii (\textit{e.g.} $>2.5\AA$)
where enhanced sampling methods are needed for proper convergence\cite{Xi2015}.  

\begin{figure}
\includegraphics{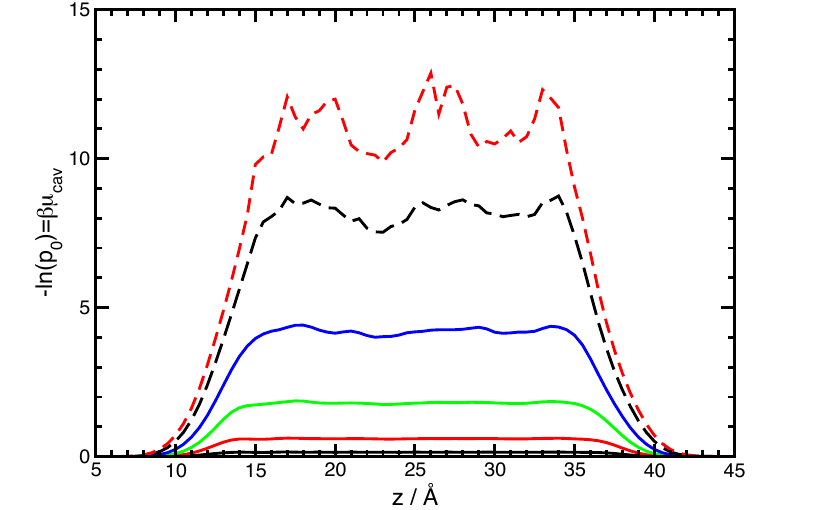}
\caption{\label{cavEnfullslabfig}  
Cavity formation free energy with revPBE-D3 (MOLOPT) for several cavity sizes (1~\AA~black solid, 
1.5~\AA~red solid, 2.0~\AA~green solid, 2.5~\AA~blue solid, 3.0~\AA~black dashed, and 3.5~\AA~red dashed line) as a function of position within the slab. This is equivalent to the potential of mean force for a hard sphere crossing the air-water interface. 
}
\end{figure}

Finally, Figure~\ref{cavEnfullslabfig} depicts the changes in the cavity formation energy as a function of $z$ for 
the revPBE-D3/MOLOPT case in the slab geometry. This is equivalent to the potential of mean force on moving a hard sphere 
solute across the air-water interface. This is an important quantity for building improved simple 
models of the distribution of solutes at the air-water
interface.\cite{Levin2009,Duignan2014b,Vaikuntanathan_pnas_16} Again, 
to probe larger cavity sizes it will be necessary to use a biasing potential to improve the sampling.\cite{Xi2015}
But overall, there is good agreement between all  models indicating  that
the molecular scale response is robust to all methods studied here.

\section{Conclusion}
\label{sec:conc}
In conclusion, we have established the simulation protocol for DFT calculations 
using popular GGA functionals
to accurately study the structure and mass density fluctuations of water across
different scales.
Moreover, we have demonstrated the quality of the structure and mass density
fluctuations is robust  across a variety of ensembles
for liquid water at 300K using DFT. 
Specifically,  the simulated structure
of water obtained from the NpT ensembles was shown to be consistent with that obtained from
the NVT ensembles in both bulk and slab simulation geometries for all DFT functionals. 
The computed density for all DFT functionals was in the neighborhood of
1 g/cm$^3$ with a slight dependence on protocol.  
Our research
suggests that revPBE-D3 provides an excellent description of water at ambient
conditions in agreement with a recent study that used a sophisticated fitting
scheme to derive an empirical potential based on revPBE-D3.\cite{Morawietz_pnas_6}
However, it was recently shown that revPBE-D3 water
with the inclusion of NQE was found to \textit{worsen} the agreement with
a variety of experiments.~\cite{Marsalek-2017} Interestingly, using the more accurate hybrid density functionals
in conjunction with NQE provide an excellent agreement with structural,
dynamic,
and spectral properties of bulk liquid water.~\cite{Marsalek-2017}
This finding is consistent with a recent calculation of water clusters up to pentamer
revealing  that revPBE-D3 benefits from a subtle cancellation of error and
that more accurate meta-GGA functionals in conjunction with NQE will provide
the correct description of liquid water.~\cite{MHG-2017}

In order to ascertain differences between the empirical  and the
DFT-based interaction potentials for water,  we have investigated the 
local structure of ambient water by looking into the distribution of the
$d_5$ order parameter that represents the distance of the 5$^{\rm th}$ nearest
neighbor from a tagged water molecule. We demonstrated that this 
order parameter probes the local heterogeneity of 
water and demonstrates that DFT based potentials exhibit a  broad range
of local environments, in contrast to the empirical models.

In addition to the mass density and local structure we examined  mass density fluctuations
and the response to molecular scale and
macroscopic (\textit e.g. air-water) interfaces.  All empirical models studied produce an 
isothermal compressibility in agreement with the experimental results. 
However, the DFT results showed rather large discrepancies depending
on the simulation protocol used in this study.
Interestingly,  when the free energy of forming a molecular sized cavity in water
was computed there was striking agreement between all representations
of interaction. It is interesting to see such good agreement in a free energy
when stark differences are present in both local structure and compressibility
between the quantum and classical representations of interaction.  
Examining the response of water to the  air-water interface also
produced excellent agreement between all representations of interaction for
the system sizes studied herein.  
Specifically, for the system sizes studied, all three
models(revPBE,MB-pol,SPC/E) were found to qualitatively reproduce 
the experimental surface tension within the framework of capillary wave theory. 

\begin{acknowledgements}  
We gratefully acknowledge John Fulton and Sotiris Xantheas for useful discussions and Prof.
Francesco Paesani for his guidance regarding the studies using MB-pol.
We also gratefully acknowledge Prof. Joost VandeVondele for discussions regarding the system setup.
This work was supported by the U.S. Department of Energy (DOE), Office of Science, Office of Basic Energy 
Sciences Grant No. BES DE-FG02-09ER46650, which supported MD simulations, 
data analysis, and manuscript preparation. PNNL is a multiprogram national laboratory operated by Battelle for 
the U.S. Department of Energy. This research used resources of the National Energy Research Scientific Computing
Center, a DOE Office of Science User Facility 
under Contract No. DE-AC02-05CH11231. Additional computing resources were generously allocated by PNNL's Institutional 
Computing program.   
\end{acknowledgements} 

\appendix
\section{Convergence studies}
\begin{figure}
\includegraphics{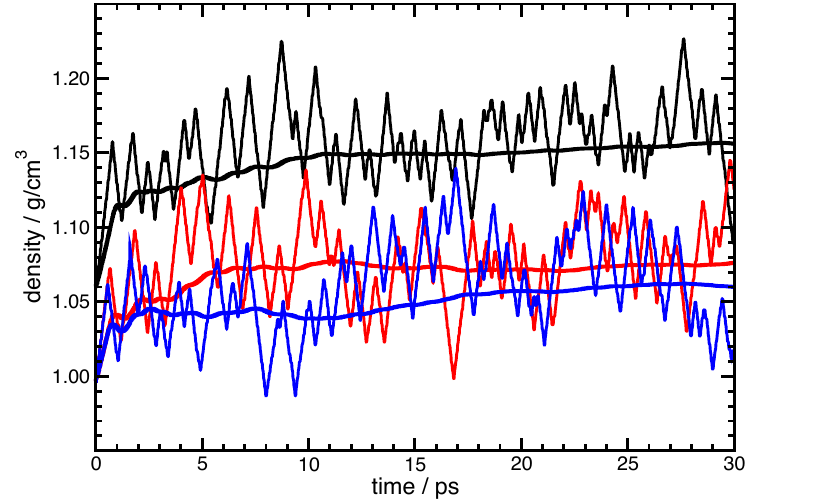}
  \caption{\label{fig:1}The instantaneous density fluctuation and its running average as a function of simulation time from 
 the NpT simulation at various density cutoff (1000 Ry black, 2000 Ry red, and 3000 Ry blue) at the revPBE-D3/TZV2P level 
 of theory with grid interpolation.}
\label{fig:1}
\end{figure}

Here we examine the effects of the electron density cutoff in conjunction with
the use of grid interpolation
techniques that are present in the CP2K code. 
We performed simulations using the 64 water box with a cutoff of 1000, 2000 and 3000 Ry 
for 30 ps each.  We  use the grid interpolation method
\textit{i.e.} the electron density is calculated using a smoothing
protocol\cite{Vande_cpc_05} using the keyword options XC\_SMOOTH\_RHO NN10  and XC\_DERIV SPLINE2\_SMOOTH.   
Figure \ref{fig:1} shows the instantaneous density and the running average with simulation time
using the aforementioned cutoffs and grid interpolation technique. All
simulations were energy conserving and produce a good liquid structure, 
however both the quality of the density fluctuations and the slow convergence
of the mass density on 
the planewave cutoff is observed in contrast to simulations highlighted in
Figure~\ref{fig:3}.
From the examination of Figure~\ref{fig:1} it is clear that the mass density
has not satisfactorily  converged even at 3000 Ry.

\begin{figure}
\includegraphics{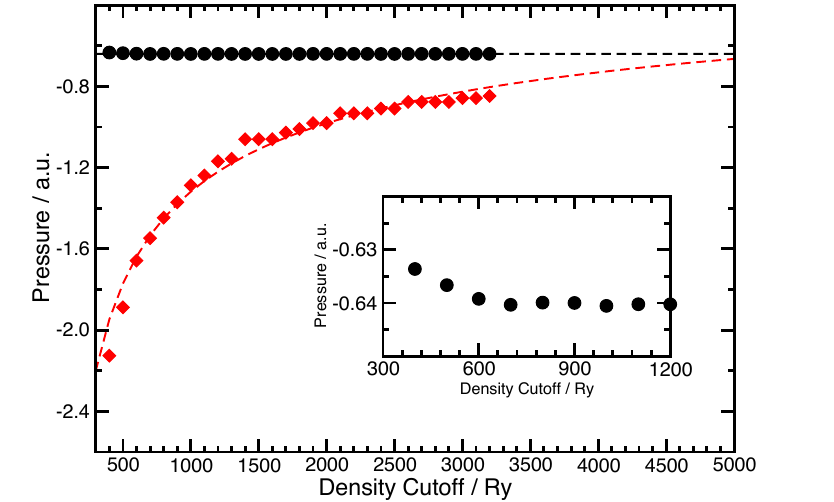}
  \caption{\label{fig:2}
The effect of density cutoff on pressure for Fourier (black circles) and grid (red diamonds) 
interpolation methods from the NpT simulation at the revPBE-D3/TZV2P level of theory. 
The inset shows the low cutoff region for Fourier interpolation.
}
\end{figure}

This can be understood by examining Figure \ref{fig:2} 
that displays  the variation of the pressure at different electron density
cutoff using both the smoothing and Fourier interpolation techniques. 
It is clear that
convergence of the pressure is achieved at $\sim$ 800 Ry when the Fourier
interpolation, namely using no smoothing protocol, is used. 
Because grid interpolation requires an abnormally high  electron density cutoff 
we choose to perform
all simulations  using the Fourier interpolation.  This affords a set of
reproducible results as a function of system size and across
ensembles where we can confidently focus on the quality of fluctuations that
are important to ascertain differences between descriptions of molecular
interaction.

\section{Functional and basis set dependence on cavity free energies }

\begin{figure}
\includegraphics{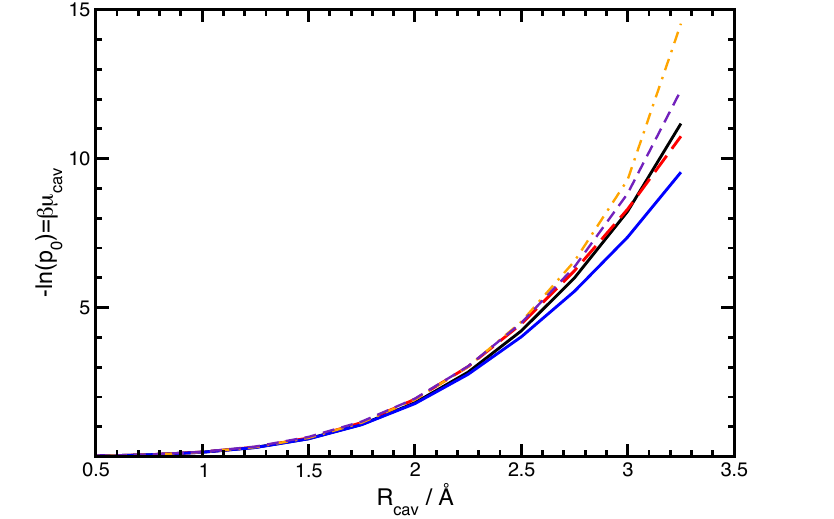}
  \caption{\label{cavEnFullfrDFTfig}
Cavity formation free energy for the different DFT water models as a function of cavity size. 
The NpT revPBE-D3/MOLOPT (red dashed) revPBE-D3/TZV2P (blue) and BLYP-D2/TZV2P (violet dashed)  and slab revPBE-D3/MOLOPT (Black) 
calculations agree. There is a non-trivial basis set and 
functional dependence for the larger cavity sizes. The NVT (yellow dash-dot) calculation is substantially too high.
  }
\end{figure}

Figure~\ref{cavEnFullfrDFTfig} is a comparison between different DFT functionals. We can see that 
the MOLOPT in the NpT ensemble calculation agrees with the MOLOPT slab calculation. 
This is to be expected as both of these simulations allow the water cell to fluctuate in 
size and are at their natural density. 
On the other hand, the NVT 
calculation  has significantly higher cavity formation energies as the
simulation cell can
not fluctuate to compensate for the cavity. The BLYP-D2 and revPBE-D3 using
TZV2P basis set results 
show some differences compared to the revPBE-D3 MOLOPT results for the larger cavity sizes indicating
that there is a degree of basis set and functional dependence to this quantity.
This study suggests that for cavities with a small radius ($\le 2\AA$) 
the NVT ensemble produces converged cavity free energies.

%
%

%


\nocite{*}
\bibliography{Journals,bib}

\end{document}